\documentstyle[12pt]{article}

\begin{document}
\title{Bound states of singlet quarks at LHC}
\author{N.V.Krasnikov \thanks{E-mail address: KRASNIKO@MS2.INR.AC.RU}
\\Institute for Nuclear Research\\
Moscow 117312, Russia\\ and CERN, Geneva}
\date{May 1996}
\maketitle
\begin{abstract}
We discuss the discovery potential of the bound states of singlet quarks  
at LHC. We find that it is possible to discover bound states of singlet 
quarks at LHC with singlet quark masses up to 300 Gev for 
$e_{Q} = \frac{2}{3}$ and up to 200 Gev for $e_{Q} = -\frac{1}{3}$.

\end{abstract}

\begin{flushleft}
CMS Technical Note TN/96-055
\end{flushleft}
\newpage

The aim of this paper is the discussion of the discovery potential of  
the bound states of singlet quarks \cite{1}-\cite{14} at LHC. Singlet 
quarks are color-triplet fermions whose left and right chiral components 
are both singlets with respect to the $SU(2)$ weak isospin gauge group. 
They can mix with the ordinary quarks and as a result of the mixing 
they decay into ordinary quarks. For small mixing $\epsilon \leq O(0.1)$ 
the lifetime of singlet quark is less than $O(10)$ Mev so singlet quarks can 
form bound states. Note that as a result of mixing of singlet quarks 
with ordinary quarks tree level flavor-changing neutral currents are 
generated. Bounds from flavor changing neutral currents typically 
restrict the mixing parameter to be smaller than $O(0.1)$. So if 
singlet quarks exist they have to be rather longlived and have to 
form bound states. The phenomenology of bound states of new quarks has been 
discussed in refs. \cite{11,12}. Recently it has been conjectured 
\cite{13,14} that singlet quarks can solve $R_{b}$ - $R_{c}$ problem. 

In this paper we study a production and decays of $0^{-+}$ and 
$1^{--}$ superheavy quarkonium states. Other superheavy quarkonium 
states $0^{++}, 1^{++}, 2^{++}$ have much smaller cross sections and 
they are practically unobservable \cite{11,12}. The main difference 
between our paper and refs. \cite{11,12} is that we use the LHC total energy 
$\sqrt{s} = 14$ Tev and STEQ2L parton distributions. Besides 
refs. \cite{11,12} consider the case of the 4 th generation doublet. 
Also we calculate the main backgrounds. The cross section 
for pseudoscalar $0^{-+}$ quarkonium production by two gluon fusion 
in $pp$ collisions can be written in terms of gluonic decay width of 
$0^{-+}$ and the gluon distribution $g(x,\mu)$ in a proton 
as \cite{11,15}
    
$$
\sigma (pp \rightarrow gg \rightarrow 0^{-+}) = 
$$
$$
\frac{\pi^2\tau\Gamma(0^{-+} \rightarrow gg)} 
{8 M^{3}_{0^{-+}}}\int^{1}_{\tau}
g(x,\mu)g(x^{-1}\tau , \mu)  
x^{-1}dx \,
\eqno{(12)}
$$
with $\tau = \frac{M^{2}_{0^{-+}}}{s}$. In nonrelativistic model the decay 
width of pseudoscalar quarkonium into two gluons is given by the  
formula \cite{1,14}
\begin{equation}
\Gamma(0^{-+} \rightarrow gg) = \frac{8\alpha_{s}^{2}
(M^2_{0^{-+}})}{3M^2_{0^{-+}}}
|R_{S}(0)|^2
\end{equation} 
Here $R_{S}(0)$ denotes the radial wave function of the S state at $r = 0$. 
For the Coulomb potential we have $ |R_{S}(0)|^{2} = 4(\frac{2}{3}
\alpha_{s}(M^{2}_{0^{-+}})M_{0^{-+}})^3$. For other phenomenological 
potentials (Cornell potential, Richardson potential, Wisconsin potential) 
the value of the radial wave function at zero radius is higher \cite{1,11} 
so the use of the Coulomb potential gives the lowest cross section 
of $0^{-+}$ pseudoscalar quarkonium production. In this paper we shall 
use the Coulomb potential. The best way to look for the pseudoscalar 
quarkonium is through its decay into two photons. The reaction 
$pp \rightarrow 0^{-+} \rightarrow \gamma\gamma$ is very similar to the 
famous reaction $pp \rightarrow Higgs \rightarrow \gamma\gamma$ which is 
supposed to be the main reaction for the Higgs boson discovery at LHC. 
The decay width of pseudoscalar quarkonium into 
two photons is 
\begin{equation}
\Gamma(0^{-+} \rightarrow \gamma \gamma) =  
\frac{12\alpha^2 e_{Q}^{4}}{M^2}|R_{S}(0)|^2
\end{equation}   
Here $e_{Q}$ is the charge of the singlet quark Q in units of the proton 
charge. 
We have calculated the cross sections for the pseudoscalar quarkonium 
production using the parton distributions STEQ2L of ref.\cite{16}.  
In our calculations we have used PYTHIA 5.7 and JETSET 7.4 generators 
\cite{17}.  We took the value of the renormalization point $\mu$ equal to 
the mass $M_{0^{-+}}$ of the pseudoscalar quarkonium. We put strong 
coupling constant $\alpha_{s}(M^2_Z) = 0.120$, effective electromagnetic 
coupling constant $\alpha \equiv \alpha(M^2_Z) = \frac{1}{128}$ and top 
quark mass $m_{t} = 175$ Gev. We have checked also that the variation  
of the renormalization point $\mu$ in the interval 
$0.5M_{0^{-+}} - 2M_{0^{-+}}$ leads to the variation of the cross sections 
less than 30 percent. In our estimates we assumed the integral luminosity 
$\int L = 10^{5}(pb)^{-1}$ and the total energy $\sqrt{s} = 14\:Tev$. 
The main background for the reaction 
$pp \rightarrow 0^{-+} \rightarrow \gamma \gamma$ comes from:

1. prompt diphoton production from quark annihilation and gluon fusion 
diagrams and bremsstrahlung from the outgoing quark line in the QCD 
Compton diagrams

2. background from jets, where an electromagnetic energy deposit originates 
from the decay of neutral hadrons in a jet or from 1 jet + 1 prompt photon. 

The jet background is reduced \cite{18} by imposing an isolation cut, which 
also reduces the bremsstrahlung background. The photon is defined an isolated 
\cite{18} if there is no charged track or electromagnetic  shower with a 
transverse momentum greater than 2.5 Gev within a region $\Delta R \geq 0.3$ 
around it. It is assumed that the jet background is reduced to an 
insignificant level ($\leq 10$ percent) by the combination of isolation and  
$\pi^{0}$ rejection cuts \cite{18}. For CMS detector \cite{18} an efficiency 
of 64 percent was assumed for reconstruction of each photon 
(i.e. 41 percent per event). 
We assumed that the accuracy of the restoration of the diphoton invariant 
mass is 1 percent that is conservative estimate for CMS detector where the 
photon energy resolution is assumed to be $\Delta E /E = 0.02/E^{0.5} \oplus 
0.005 \oplus 0.2/E$ in the barrel and $\Delta E/E =0.05/E^{0.5} \oplus 0.005 
\oplus 0.2/E$ in the endcap, where there is a preshower detector. In our 
calculations we have used the following cuts for photons:
$$
|\eta| \leq 2.5, \, \, \, |p_{t}| \geq 25 Gev
$$
The results of our calculations are presented in table 1. Our main 
conclusion is that at LHC for integral luminosity $L = 10^{5}$ $pb^{-1}$ it 
would be possible to discover the pseudoscalar quarkonium 
at the $5\sigma$ level for $e_{q} = -\frac{1}{3}$ with singlet quark 
masses $m_{Q}$ up to 100 Gev and for $e_{Q}= \frac{2}{3}$ with singlet 
quark masses up to 300 Gev. 

For the estimate of the vector quarkonium $1^{--}$ production we use the 
cross section for the subprocess $gg \rightarrow 1^{--} g$ equal to 
\cite{11,12}
\begin{equation} 
\sigma(gg \rightarrow 1^{--} g) = \frac{9\pi^2}{8M^3_{1^{--}}(\pi^2-9)}\Gamma
(1^{--} \rightarrow ggg)I(\frac{S}{M^2_{1^{--}}})  \,,
\end{equation}
where
\begin{equation}
I(x) = \frac{2}{x^2}(\frac{x+1}{x-1} - \frac{2xln(x)}{(x-1)^2}) + 
\frac{2(x-1)}{x(x+1)^2} + \frac{4ln(x)}{(x+1)^3}   
\end{equation}
The decay width of vector quarkonium into 3 gluons is determined by the 
formula \cite{1,12}
\begin{equation}
\Gamma(1^{--} \rightarrow ggg) = \frac{40(\pi^2-9)\alpha^3_s}{81\pi 
M^3_{1^{--}}}|R_{S}(0)|^2
\end{equation}   
The most promising decay mode for the detection of vector quarkonium is 
$ 1^{--} \rightarrow \mu^{+}\mu^{-}$ or $1^{--} \rightarrow e^{+}e^{-}$. 
The main background for the process 
$pp \rightarrow 1^{--} +... \rightarrow \mu^{+}\mu^{-} +...$ is the 
Drell-Yan process. We have calculated the corresponding cross sections 
for singlet quarks with electric charge $E_{Q} = -\frac{1}{3} $ 
(analog of $d_{R}$-quark) and for singlet quark with electric charge 
$e_{Q} = \frac{2}{3}$ (analog of $u_{R}$-quark). The results of our 
calculations are presented in table 2. Our main conclusion is that for 
$e_{Q} = -\frac{1}{3}$ it would be possible to discover singlet quarks 
with the mass up to 200 Gev and for $e_{Q} = \frac{2}{3}$ it would be 
possible to discover singlet quarks with the mass up to 250 Gev.

It should be noted that an account of the production of the 
radial excitations of $0^{-+}$ and $1^{--}$ quarkoniums leads to the 
increase of the signal cross section approximately by 20 percent 
\cite{11}.  
  
To conclude, in this note we have studied the perspectives of the discovery
of the bound states of singlet quarks, namely pseudoscalar quarkonium  
$0^{-+}$ and vector quarkonium $1^{--}$ at LHC. The best way to 
detect such bound states  is the measurement of the diphoton and dilepton 
invariant masses. LHC will be able to discover the  superheavy 
quarkoniums  with singlet quark masses up to 300 Gev for $e_{Q} = \frac{2}{3}$ 
and and up to 200 Gev for $e_{Q} =-\frac{1}{3}$.      
    
I am indebted to the collaborators of the INR theoretical department for 
discussions and critical comments. The research described in this publication 
was made possible in part by JSPS program on Japan-FSU Scientists 
Collaboration.

\newpage

Table 1. The cross sections and branchings for the process 
$pp \rightarrow 0^{-+} +...\rightarrow \gamma \gamma +...$. Here:

1. $\sigma \equiv \sigma(pp \rightarrow 0^{-+} \rightarrow gg)
\cdot \Gamma^{-1}(0^{-+} \rightarrow gg)$ in $pb \cdot(Mev)^{-1}$.

2. $\sigma_{cut} \equiv \sigma(pp \rightarrow 0^{-+} \rightarrow gg 
||\eta_{g}| \geq 2.5, |p_{tr,g}| \geq 25 Gev) \cdot \Gamma^{-1}(0^{-+} 
\rightarrow gg)$ in $pb \cdot (Mev)^{-1}$ .

3. $\Gamma \equiv  \Gamma (0^{-+} \rightarrow gg)$ in Mev.

4. $\sigma_{\gamma,cut} \equiv \sigma(pp \rightarrow 0^{-+} \rightarrow 
  \gamma \gamma ||\eta_{\gamma}| \leq 2.5, |p_{tr,\gamma}| \geq 25 Gev)$ in
  $fb \cdot 10^{-1}$ for $e_{Q} = -\frac{1}{3}$.

5. $\sigma_{Back} \equiv d\sigma_{Back}(|\eta_{\gamma}| \leq 2.5, 
|p_{tr, \gamma}| 
\geq 25 Gev)/dm_{\gamma\gamma}\cdot10^{-2}M_{0^{-+}}$ 
in $fb$.

6. $L_{1} \equiv k \cdot \frac{N_{S}}{\sqrt{N_{Back}}}$, 
 $N_{S} \equiv \sigma_{\gamma, cut}\cdot L$, $N_{Back} \equiv \sigma_{Back}
\cdot L$,  $L = 10^{5}\cdot pb^{-1}$, $k = 0.64$, $e_{Q} = -\frac{1}{3}$.

7. $L_{2} \equiv \frac{N_{S}}{\sqrt{N_{Back}}}$, $N_{S} \equiv 
16k \cdot \sigma_{\gamma, cut}\cdot L$, $N_{BacK} \equiv \sigma_{Back} \cdot L$,  
$L = 10^{5}\cdot pb^{-1}$, $k = 0.64$, $e_{Q} = \frac{2}{3}$   

8. $M_{Q}$ is the mass of the singlet quark Q.

\begin{center}
\begin{tabular}{|l|l|l|l|l|l|l|l|}
\hline
$M_{Q}$(Gev) &$\sigma$&$\sigma_{cut}$&$\Gamma$&$\sigma_{\gamma,cut}$&$\sigma_{Back}$&$L_1$&$L_2$  \\ 
\hline
100 & 22.4 & 16.6  & 1.8 & 75 & 64 & 6.0 & 96   \\
\hline
125 & 7.3 & 5.6 & 2.0 & 29 & 35 & 3.2 & 50 \\
\hline
150 & 3.9 & 3.1 & 2.1 & 18 & 21  & 2.6 & 41  \\
\hline
175 & 1.9 & 1.6 &2.3 & 10.7 & 12 &2.0 & 31 \\
\hline
200 & 1.0 & 0.83 & 2.3 & 5.7 & 8.4 & 1.3 & 20 \\
\hline
225 & 0.40 & 0.34 & 2.4 & 2.7 & 6.3 & 0.69 & 12 \\
\hline
250 & 0.31 & 0.27 & 2.5 & 2.2 & 4.4 & 0.67 & 11 \\
\hline
275 & 0.19 & 0.17 & 2.6 & 1.4 & 3.4 & 0.50 & 8.0 \\
\hline
300 & 0.12 & 0.11 & 2.7 & 0.98 & 2.5 & 0.39 & 6.3 \\
\hline
350 & 0.053 & 0.047 & 2.8 & 0.46 & 1.5 & 0.25 & 3.9 \\  
\hline
400 & 0.026 & 0.023 & 3 & 0.25 & 0.96 & 0.16 & 2.6 \\
\hline

\end{tabular}
\end{center}

\newpage

Table 2. The cross sections and branchings for the process 
$pp \rightarrow 1^{--} + ... \rightarrow \mu^{+}\mu^{-} + ...$. Here:

1.  $\sigma_{cut,1} \equiv \sigma(pp \rightarrow 1^{--} +... \rightarrow 
\mu^{+}\mu^{-} + ...| |\eta_{\mu}| \leq 2.5, |p_{tr,\mu}| \geq 5 Gev)$ 
in fb for $Q = -\frac{1}{3}$.

2.  $\sigma_{cut,2} \equiv \sigma (pp \rightarrow 1^{--} + ... \rightarrow 
\mu^{+}\mu^{-} + ...| |\eta_{\mu}| \leq 2.5, |p_{tr,\mu}| \geq 5 Gev$ in 
fb for $e_{Q} = \frac {2}{3}$.

3. $\sigma_{Back} \equiv d\sigma_{DY}(pp \rightarrow \mu^{+}\mu^{-} +...|
|\eta_{\mu}| \leq 2.5, |p_{tr,\mu}| \geq 5 Gev)/dm_{\mu\mu} \cdot M_{1^{--}}
\cdot (2\cdot 10^{-2})$. 
 
4. $L_{1} \equiv \frac{N_{S}}{\sqrt{N_{Back}}}$, $N_{S} \equiv \sigma_{cut,1}
\cdot L$, $N_{Back} \equiv \sigma_{Back}\cdot L$, $L = 10^{5} pb^{-1}$,
$e_{Q} = -\frac{1}{3}$.

5. $L_{2} \equiv \frac{N_{S}}{\sqrt{N_{Back}}}$, $N_{S} \equiv \sigma_{cut,2}
\cdot L$, $N_{Back} \equiv \sigma_{Back}\cdot L$, $L = 10^{5} pb^{-1}$, 
$e_{Q} = \frac{2}{3}$.

6. $M_{Q}$ is the mass of singlet quark Q.

\begin{center}
\begin{tabular}{|l|l|l|l|l|l|}
\hline
$M_{Q}$(Gev)&$\sigma_{cut,1}$& $\sigma_{cut,2}$&$\sigma_{Back}$&$L_{1}$&$L_{2}$\\
\hline
100 & 48 & 80 & 176 & 36 & 60 \\
\hline
125 & 17 & 29 & 78 & 19 & 33 \\
\hline
150 & 10 & 17 & 44 & 15 & 26 \\
\hline
175 & 5.1 & 8.5 & 26 & 10 & 17\\
\hline
200 & 2.8 & 4.6 & 17 & 6.8 & 11 \\
\hline 
225 & 1.1 & 1.9 & 12 & 3.1 & 5.4 \\
\hline
250 & 0.91 & 1.5 & 8 & 2.8 & 5.1 \\
\hline

\end{tabular}
\end{center}

\end{document}